# Improved annotation of 3' untranslated regions and complex loci by combination of strand-specific Direct RNA Sequencing, RNA-seq and ESTs

Nick Schurch[1,5,6,*], Christian Cole[1,5,6,*], Alexander Sherstnev[1], Junfang Song[2], Céline Duc[4,], Kate G. Storey[2], W. H. Irwin McLean[3], Sara J. Brown[3], Gordon G. Simpson[4,7], and Geoffrey J. Barton[1,5,6,+]

[1]Division of Computational Biology, [2]Division of Cell and Developmental Biology; [3]Division of Molecular Medicine, [4]Division of Plant Sciences, [5]Division of Biological Chemistry and Drug Discovery, [6]Centre for Gene Regulation and Expression,

College of Life Sciences, University of Dundee, Dow Street, Dundee, DD1 5EH, UK

[7]Cell and Molecular Sciences,
The James Hutton Institute, Invergowrie, Dundee DD2 5DA, UK

* joint first authors
+Corresponding author

# Abstract

The reference annotations made for a genome sequence provide the framework for all subsequent analyses of the genome. Correct annotation is particularly important when interpreting the results of RNA-seq experiments where short sequence reads are mapped against the genome and assigned to genes according to the annotation. Inconsistencies in annotations between the reference and the experimental system can lead to incorrect interpretation of the effect on RNA expression of an experimental treatment or mutation in the system under study. Until recently, the genome-wide annotation of 3' untranslated regions received less attention than coding regions and the delineation of intron/exon boundaries. In this paper, data produced for samples in Human, Chicken and *A. thaliana* by the novel single-molecule, strand-specific, Direct RNA Sequencing technology from Helicos Biosciences which locates 3' polyadenylation sites to within +/- 2 nt, were combined with archival EST and RNA-Seq data. Nine examples are illustrated where this combination of data allowed: (1) gene and 3' UTR re-annotation (including extension of one 3' UTR by 5.9 kb); (2) disentangling of gene expression in complex regions; (3) clearer interpretation of small RNA expression and (4) identification of novel genes. While the specific examples displayed here may become obsolete as genome sequences and their annotations are refined, the principles laid out in this paper will be of general use both to those annotating genomes and those seeking to interpret existing publically available annotations in the context of their own experimental data.

# Introduction

The majority of applications of a genome sequence rely on the gene structures and associated features provided by the reference genome annotation. Methods to annotate a newly sequenced genome are well developed and exploit both data-driven and *ab initio* feature prediction [1, 2], but annotation is always derived from a snapshot of knowledge at the time it is carried out. As new data become available, the annotation must be revised if it is to remain relevant and useful (e.g. [3-6]). Annotation projects for the most complete and well described metazoan genomes: human[7]; mouse[8] and zebrafish[9], combine automatic methods with manual curation to provide an authoritative annotation that is regularly updated by incorporating new experimental data (e.g. [10]). The reference annotations for most other genomes rely more heavily on fully automatic annotation with limited manual curation. Since the structure of the gene transcript can vary according to cell type, treatment and other stimuli, the annotation that is most relevant may need to be re-defined for each set of experimental conditions. Advances in short-read, high-throughput transcript sequencing (RNA-seq) and its use in differential expression analysis have highlighted the importance of accurate gene models and prompted the development of methods to carry out experiment-specific predictions of gene structure (e.g. see [2, 11-14]). However, conventional RNA-seq experiments often do not define the ends of genes with high precision. Incorrect assignment of the 5' and 3' UTRs may cause reads in an RNA-seq experiment to be assigned to intergenic regions and so give erroneous estimates of gene expression. Furthermore, the short read length may not provide evidence for an unambiguous gene structure where there are overlapping genes, while RNAseq data that are not strand-specific are complex to apply in areas where genes overlap.

Recently, techniques have been developed that allow sites of cleavage and polyadenylation at the 3'-end of transcripts to be identified in a high-throughput manner. These include 3P-Seq which has been applied to the characterisation of 3'UTRs in *C.elegans* [15] and zebrafish [16] and Helicos Bioscience's single-molecule direct RNA sequencing (DRS) [17] which has been applied to large-scale 3'UTR studies in human [18] *A. thaliana* [19], and yeasts [20, 21] DRS [17] captures RNA by the poly(A) tail and sequences the RNA immediately adjacent, so giving a very clear read-out of the transcript's 3'-end. DRS is strand-specific, has no amplification step, is less susceptible to internal priming than other methods and since it sequences RNA not DNA, does not require reverse transcription and the artefacts that can generate .

DRS has been used in an automatic protocol to re-annotate the 3'-ends of over 10,000 protein coding genes in *A. thaliana* of which more than 3,400 were extended by at least 10 nt. [19]. Here, this study is extended to explore the potential of combining DRS with conventional RNA-seq, small RNA-seq (sRNA-seq) and archival expressed sequence tag (EST) data for genome annotation in human, chicken and *A. thaliana*. Combining DRS, RNA-seq, EST and sRNA-seq data promises to mitigate the limitations of each individual technology; providing multiple, orthogonal, sources of evidence for gene intron/exon structure, 3' UTR regions and mature small RNAs and microRNAs, even in complex genomic regions.

# Materials and Methods

The collection of human skin samples was performed under the governance of the Tayside Tissue Bank after review and approval by the Tayside Tissue Bank Committee (ref TR000192). Skin was obtained, with written informed consent, as tissue discarded from plastic surgical procedures and in accordance with the Declaration of Helsinki. Data generated from the skin analysis were analysed anonymously.

In this paper data from the authors' own laboratories were combined with data from public archives. The source of all data presented here is described below.

## *Gallus gallus* (chicken) DRS Data

### Sample Dissection

Pre Neural Tube (hereafter PNT) explants were dissected from Hamburger and Hamilton stage 10, 10 to 12 somite chick embryos ([22]). The explant was taken from a region rostral to the node and at a two presumptive somite distance from the last somite formed (somite I). The notochord was removed by controlled trypsin digestion aiming to keep the neural ventral midline. Dissections were carried out in L15 medium at 4°C and explants were taken for RNA extraction and DRS sequencing from three individual embryos (biological replicates).

### RNA Extraction & Quality Testing

All surfaces and dissecting tools were treated with RNAZap (Ambion) and rinsed with DEPC-treated water. RNA was extracted from the three PNT explants in Trizol reagent (Invitrogen) by phase separation with chloroform, followed by precipitation with isopropanol and linear acrylamide. The RNA was washed in 70% ethanol, air-dried, re-suspended in DEPC-water and frozen in liquid nitrogen. Total RNA was quantified and quality tested using the Agilent RNA assay (Agilent Bioanalyser pico RNA chip) by Helicos Biosciences. Samples with a RIN number above 8.0 were selected, and were then sequenced by DRS ([17]), producing 7.2-16.4 million raw reads per sample.

## DRS Data Processing

Raw DRS reads from each sample were mapped to v2.1 of the chicken genome (galGal3) with Helicos Biosciences' open-source mapping pipeline *Helisphere* (v2.0.022410) with the default parameters. The mapped reads were then filtered with four additional selection criteria to remove as much noise from the data as possible. Only reads with unique, high-quality, mappings to the genome (both locally and globally) were accepted. DRS sequencing technology is prone to producing reads that require a large number of insertions or deletions (in-dels) to align to the genome ([17, 23]). Accordingly, to minimise ambiguity only reads whose best-match alignments contained fewer than four indels, and whose read length was greater than 25 bases were accepted. Finally, all reads that map to any positions in the genome with fewer than 3 reads coverage per replicate were discarded. Based on the existing chicken genome annotations from Ensembl, this resulted in a total of ~5,178 Ensembl genes with measured expression in all three PNT DRS replicate datasets. Data are available from www.compbio.dundee.ac.uk/polyadb and will be deposited at the Short Read Archive.

# Public *Gallus gallus* Illumina RNA-seq Data

The publicly available chicken Illumina RNA-seq data discussed here forms part of a study that examined gene expression in mammalian organs (Short Read Archive study: SRP007412 GSE30352 - [24]). This study used the Illumina Genome Analyser IIx platform to generate 76bp reads for six tissues (brain - cerebral cortex or whole brain without cerebellum, cerebellum, heart, kidney, liver and testis) from one male and one female per somatic tissue (two males for testis). Data for the chicken were generated for this mammalian-focussed study as an evolutionary outgroup. The data were downloaded from the Short Read Archive, converted to fastq format with the SRA toolkit (v2.1.10, http://www.ncbi.nlm.nih.gov/Traces/sra/sra.cgi?view=software). The reads in each dataset were then aligned to v2.1 of the chicken genome (galGal3) with the splice-aware alignment software *TopHat* (v2.0.0, http://tophat.cbcb.umd.edu/ - [14]) in conjunction with *Bowtie* (v2.0.0 beta5,

 - [13]), with the *--coverage-search*, *--microexon-search* and *--b2-very-sensitive* options in addition to the *TopHat* defaults. Combined, the twelve samples total ~251M reads, 64% (~161M reads) of which map to the genome using these settings.

## *Homo sapiens* skin DRS data

### Sample Dissection

A clinically normal human skin sample was obtained by 4mm punch biopsy of skin tissue removed during plastic surgical procedures from the abdomen of an adult female, with approval from the local Research Ethics Committee, under the governance of Tayside Tissue Bank. The biopsy sample was snap frozen in liquid nitrogen and stored at -80°C. The specimen was disrupted and homogenised using a TissueLyser (Qiagen TissueLyser LT, Qiagen, UK) at 50 oscillations per second for 5 minutes at 4°C. Total RNA (>200nt in length) was extracted using the Qiagen RNeasy® Mini Kit according to manufacturer's protocol and stored at -80°C prior to RNA sequencing. Sequencing was performed as previously described ([17]).

### DRS Data Processing

The raw sequence data was aligned to the GRCh37 release of the human genome with the open source HeliSphere package (version 1.1.030309). Specifically *indexDPgenomic* was run with the following parameters set: *--best_only --min_norm_score 4.0 --strands both --alignment_type GL* the remainder were kept to their defaults. Aligned data were filtered with *filterAlign* in order to return only unique alignments from reads at least 25bp in length (~7M reads remaining). Further filtering was applied with in-house scripts to remove reads with indels larger than four bases and singleton positions where only one read was found, leaving 4,974,304 DRS reads for further analysis. The data are available from www.compbio.dundee.ac.uk/polyadb and will be deposited in the Short Read Archive.

## *Homo sapiens* Illumina RNA-seq Data

A publicly available dataset was downloaded from the Short Read Archive (Accession: SRX084679). As no skin sample data was available, these data were from normal human epidermal keratinocyte (NHEK) whole cells. The polyA+ purified RNA was sequenced as 76bp paired-end reads resulting in 46.4M read pairs in sample SRR315327.

All the reads were then aligned to the GRCh37 release of the human genome with *TopHat* (v2.0.0) with the *--coverage-search*, *--microexon-search* and *--b2-very-sensitive* options set in addition to the *TopHat* defaults. Of the 46.4M read pairs, 93.3% (43.3M pairs) aligned to the genome using these settings.

## *Homo sapiens* sRNA-seq Data

Publically available data from a normal skin biopsy sample was downloaded from the Short Read Archive (Accession: SRX091761 [25]. The accession contains one sample (SRR) of ~21M 36bp single-end reads prepared via the Illumina small RNA-seq protocol. The raw reads were quality clipped, had their adapter sequences removed and any remaining reads shorter than 16bp were discarded as previously described [26]. The remaining 18,722,725 reads were collated as 788,334 unique sequences for alignment to the genome. The sequences were aligned to the GRCh37 release of the human genome with *bowtie* v0.12.3 (parameters: *-a --best --strata -v 1*).

## *Arabidopsis thaliana* DRS data

### RNA Extraction

*A. thaliana* WT Col-0 seeds were sown in MS10 plates, stratified for 2 days at 4°C and grown at a constant temperature of 24°C under 16 h light/8 h dark conditions. 14 days old seedlings were harvested. Total RNA was purified using an RNeasy kit (Qiagen). No subsequent poly(A) of the RNA was performed and further procedures in preparation or sequencing were carried out as described in [17]**.**

Raw DRS sequences were aligned by the open-source HeliSphere package (version 1.1.498.63), to the TAIR10 release of the *A. thaliana* genome. The indexDPgenomic aligner was run with seed_size=18, num_errors=1, weight=16, best_only=1, max_hit_duplication=25, percent_error=0.2; read_step=4, min_norm_score=4.2, and strands=both options. Globally non-unique alignment hits were discarded and one hit selected at rand if there were several non-unique local hits found in a genetic region. Reads with more than four indels were discarded and read alignments refined by an iterative multiple alignment procedure while DRS reads containing low complexity genomic regions, as identified by DustMasker from the Blast+ 2.2.24 package, were discarded, as previously described [19]. The data have been deposited European Nucleotide Archive (ENA): Study, PRJEB3993; accession no, ERP003245.

## *Arabidopsis thaliana* RNA-seq data

RNA-seq reads available in the accession SRR394082 were taken from the European Nucleotide Archive. These reads were generated from total RNA extracted from 10 day-old seedlings of *A. thaliana* (Columbia-0 ecotype) and sequenced by Illumina HiSeq 2000. All details of material preparation are described in [27]. The 51.8M raw reads length of 50 bp were aligned with the splice-aware alignment software *TopHat* v2.0.0 (this version of TopHat uses *Bowtie* v2.0.0 beta5) with the *--b2-very-sensitive* option in addition to the *TopHat* default options against the TAIR10 release of the *A.thaliana* genome. The total number of uniquely aligned reads was 48.8M (94.2% of the raw reads).

## *Arabidopsis thaliana* small RNA-seq data

Publicly available small RNA-seq data were taken from the European Nucleotide Archive (accession number is SRR16393810). Total RNA for these data was extracted from immature flowers of wild-type *A. thaliana* (Columbia-0 ecotype), processed with Illumina Small RNA Sample Prep Kit and sequenced with HiSeq 2000 (Illimuna). The RNA extraction and sequencing procedures are described in detail in [28]. The accession consists of 34.2M of 36 bp non-aligned reads. The raw reads were quality-clipped, had their adapter sequences removed and remaining reads shorter than 16bp were discarded as previously described [26]. The remaining 13M reads were collated as 500,000 unique sequences for alignment to the genome. The sequences were aligned to the TAIR10 release of the *A. thaliana* genome with *bowtie* v0.12.3 (parameters: *-a --best --strata -v 1*).

### *Arabidopsis thaliana* EST data

The *A. thaliana* EST data available in IGB were taken from the PlantGDB resource which aggregates the EST sequences from GenBank's nucleotide database and splits them by species. The sequences used here are from GenBank version 187. They can be downloaded in fasta format from ftp://ftp.plantgdb.org/download/FASTA_187/EST/Arabidopsis_thaliana.mRNA.EST.fasta

# Results

In this work, the definitions of 'gene' and 'gene-associated regions' (GARs) as suggested by Gerstein and colleagues [29] are followed. The results are divided into four sections where the major strengths of combining DRS data with other high-throughput transcriptomics data are highlighted by nine examples of feature re-annotation of genes and their GARs. Section 1 focusses on how the broad-coverage of RNA-seq and EST data help to bridge the gap between existing annotations and the DRS read data, enabling improved annotation of transcribed, polyadenylated regions. Section 2 illustrates how the positional specificity and native stranded-ness of DRS data enable re-annotation of complex genomic regions, without which the RNA-seq data could not be used effectively either for re-annotation or further downstream analysis. Section 3 examines the synergy between standard RNA-seq, DRS and sRNA-seq data in providing a more complete picture of non-coding RNA expression than any of these datasets can provide individually. Section 4 briefly considers the potential for combined data to enable the discovery of new genes.

## Section 1: Gene and 3' UTR re-annotation by combining DRS and RNA-seq data

### A simple example: Chicken *BMPR1A*

The chicken genome sequence and gene models based on EST data were first released in 2004 (International Chicken Polymorphism Map[28]) with a second, more complete revision (v2.1) released 2006. A draft update to v2.1 was released in 2012, but this is yet to be annotated fully. Accordingly, most current research relies on v2.1 and its annotations and does not take account of evidence from DRS experiments.

Figure 1 shows the genomic context and information sources around *BMPR1A*, a gene important in development (*F1P3H0_CHICK, ENSGALG00000002003*; [30-32]). The annotation of this gene and its GARs differ between Ensembl and RefSeq. Ensembl presents a single gene model and two short novel protein coding models. The canonical transcript (*ENSGALT00000003119*, see Table 1) covers 39,530 bp with twelve exons of 100298 bp, and an associated 228 bp 3' UTR. In contrast, the RefSeq annotation covers 39,340 bp, including a 21 bp longer first exon and a 17 bp shorter 3' UTR. Although the basic gene intron/exon structure and the 5' UTR are annotated in Ensembl/RefSeq, no 3' UTR is present in the RefSeq annotation and the 3'UTR is short in the Ensembl annotation. There is no peak in the DRS data at the end of either the RefSeq or Ensembl 3' UTR, but there are four peaks ~1.45, 1.9, 2.4 & 4.2 kb downstream of the existing Ensembl annotation (Figure 1, Track A, 1-4, respectively). These peaks all have canonical AATAAA poly(A) motifs (≤1 mismatch) located 15-22 bp upstream suggesting they are genuine poly(A) sites, however the DRS data alone do not reveal which, if any, of these sites should be associated with *BMPR1A*.

EST and RNA-seq data can provide a bridge between the Ensembl/RefSeq annotations and the DRS data. Despite their low depth, the *G. gallus* EST data show almost continuous coverage between the end of the 3' UTR annotated in Ensembl and the most 3' DRS peak. However, the EST data are not conclusive; there is a 400 bp gap in the EST coverage and the implied exon structure is inconsistent with the existing annotations. The addition of publically available RNA-seq data ([24]) strengthens the confidence that the DRS peaks correspond to the 3'-end of *BMPR1A*. The RNA-seq data cover the proposed 3' UTR and finish 1 bp beyond the fourth DRS peak. The RNA-seq data also confirm the exon/intron structure of the existing gene annotations.

Although the RNA-seq data are non-uniformly distributed, there are only three places in the proposed 3' UTR where the read depth drops to zero. In all three examples, there is good supporting evidence from overlapping ESTs that these gaps are unlikely to represent the end of the gene. The combination of DRS, EST and RNA-seq suggests the *BMPR1A* gene in *G. gallus* should be re-annotated as shown in Table 1. The new annotation indicates four alternative poly(A) sites exist in the developing chicken embryo, but there is no evidence to support the two short novel protein coding models Ensembl also provide as annotations for this gene.

## Complex, ambiguous, feature re-annotations: Chicken *HOXA7*

The re-annotation of *BMPR1* was comparatively straightforward because the different datasets reinforce each other. A more complex and ambiguous re-annotation is illustrated in Figure 2 for the *HOXA7* gene (ENSGALG00000011061, [33]). The Ensembl annotation has a single transcript that covers 1,702 bp and includes two exons (280 and 285 bp) and a short (36 bp) 3' UTR. In contrast, the RefSeq annotation covers 1,837bp, includes three exons (278, 283 & 41bp respectively) and has no defined 3' UTR.

The intron/exon structure of *HOXA7* shown in Figure 2 appears to be simpler than *BMPR1A*. However, the DRS, EST and RNA-seq datasets suggest this gene may have a more complex structure than defined in Ensembl/RefSeq. Multiple peaks are evident in the observed DRS dataset (Figure 2, Track K, 1-6) that mark potential poly(A) sites associated with *HOXA7*. The first peak (1) lies within the intron separating the two primary exons of the gene. The second peak (2) is composed of three smaller peaks that all lie within 30 bp of the end of the existing Ensembl annotation. On the surface, these appear to support the existing 3' UTR annotation, but the presence of a large peak in the DRS data 1.5kb downstream (6), if genuinely associated with *HOXA7*, suggests an alternative annotation that would not only extend the 3' UTR, but would also be the dominant transcript in the DRS dataset for this gene. Peak 6 shows a canonical AATAAA poly(A) motif 19 bp upstream, consistent with a genuine poly(A) site. Peaks 2-5 show long runs of adenosine bases immediately downstream of each peak, suggesting that they might be the result of internal priming while peak 1 shows neither of these features and it remains unclear whether it is a true site of alternative polyadenylation.

In a similar fashion to the example shown in Figure 1 (Section 1.1), both the EST and RNA-seq data bridge the gap between DRS peak 6 and the existing reference annotations. Together, these data support the proposed 3' UTR re-annotation, despite the EST data including a 500 bp region where the coverage is low (≤2 ESTs) and from an inferred exon structure that is inconsistent with the existing annotation.

While the RNA-seq data support the proposed 3' UTR re-annotation, they do not match the short initial exon present in the RefSeq annotation and the EST data. The genomic sequence in the 31 bp intron between the first and second exons in the RefSeq annotation is marked as 'N's in the genomic sequence, making it difficult to draw robust conclusions on the structure of the gene in this region. Although this exon annotation is broadly supported by the EST dataset, these data extend beyond the RefSeq annotation suggesting a potential re-annotation of the 5' UTR.

This example shows considerable non-uniformity in the RNA-seq data that map to the suggested 3' UTR, with several significant (>50 bp) gaps in the RNA-seq coverage. The EST coverage and the lack of known polyadenylation motifs in the genomic sequence surrounding these gaps suggest that these are artefacts intrinsic to the Illumina RNA-seq protocol and do not represent the end of the 3' UTR associated with *HOXA*7.

Accordingly, a re-annotation of the *HOXA7* gene in *G. gallus* (Table 2) based on the combination of DRS, EST and RNA-seq data is proposed.  The annotation broadly supports the existing intron/exon structure of the RefSeq annotation, but extends the 3' UTR by 1.5 Kb and suggests an alternative polyadenylation site. The presence of the first intron is not strongly supported by the RNA-seq data and may well be spurious or an extension of the larger second exon, or specific to a particular tissue type or biological condition not sampled by the RNA-seq experiment.

## Gene and 3' UTR re-annotation for *Homo sapiens SLFN5*

Although the human genome is actively curated, gene models can still be revised with new data. For example, *SLFN5* in *H. sapiens* until recently had a significantly truncated 3' UTR.  Prior to v69 (Oct 2012), the SLFN5 Ensembl annotation was composed of two alternative gene models; one covering 4,625 bp spanning 4 exons, and the other covering 2,540 bp spanning 3 exons. The RefSeq annotation contained a single gene model covering 4,654 bp spanning 4 exons. All these annotations included a short 5' UTR encompassing a long intron and a well-defined 1.8 kb 3' UTR. In the v69 Ensembl release, the annotations for *SLFN5* changed considerably. The 3' UTR for the primary transcript was extended by ~6kb and a third, shorter gene model was added. To date (Feb 2013), there has been no change in the RefSeq annotation for this gene. Figure 3 shows the genomic context around *SLFN5* with the most recent annotations from Ensembl and RefSeq. Both the DRS and RNA-seq data show evidence for transcription continuing up to ~6 kb further downstream than the current RefSeq annotation, and in agreement with the current Ensembl annotation. However, the DRS data reveals two alternative polyadenylation sites ~5 kb and ~8.5 kb (Figure 3, Track A, 1-2, respectively) from the first stop codon in *SLFN5*, both of which have the canonical AATAAA cleavage and polyadenylation signal upstream (19 & 24 bases, respectively) of the DRS peak. One of these sites is coincident with the Ensembl gene model, but the second site suggests a fourth alternative gene model. The combination of the DRS and RNA-seq data suggests the *SLFN5* gene in *H. sapiens* should be re-annotated as described in Table 3.

## Extension of 3' UTR for *A. thaliana*: AT4G02715

The genome of *A. thaliana* has been extensively studied since it was sequenced and released in 2000 ([34]). However, examination of the first DRS data for *A. thaliana* [19] enabled the 3'-ends of ~65% of its genes to be re-annotated automatically by considering reads within 300 bp of the TAIR10 annotated 3'-end. Sherstnev *et al* [19] only considered DRS data and this approach missed further re-annotation possibilities. For example, Figure 4 summarises the region around *AT4G02715*. The TAIR 10 annotation for this gene consists of a 0.6 kb 5'-UTR containing a single intron followed by a single 0.6kb exon. No significant DRS peaks are found within the 300bp window downstream of the 3' end of the current annotation and so the algorithm described in [19] did not re-annotate the 3' end of this gene. A cluster of DRS signals is observed ~0.6kb downstream (Figure 4, Track K, 2) followed by a set of peaks ~0.65kb further downstream (Figure 4, Track K, 3) and another cluster of peaks ~0.25 kb still further downstream (Figure 4, Track K). The RNA-seq data covers the full extent of the downstream region up to DRS peak 3. Like many poly(A) sites in *Arabidopsis*, peak 3 is composed of at least four peaks of varying strength, several of which are broader than the ±2bp positional accuracy of the DRS data [19]. The RNA-seq data also identify an intron ~1kb upstream of the end of the current annotation. The protein coded by *AT4G02715* has yet to be characterized and the current annotation represents the longest ORF in this genomic region, suggesting that the proposed extension reflects the 3' UTR of this gene. The RNA-seq data show weak expression extending out to within a few bases of peak 4, but the unmatched nature of the DRS and RNA-seq samples makes it difficult to draw strong conclusions about the nature of this region. It is possible this region is an alternative transcript for *AT4G02715* that is not expressed in the archival RNA-seq dataset.

Table 4 shows the proposed re-annotation of *AT4G02715* in *A. thaliana* based on the RNA-seq and DRS data. In the new annotation, the DRS data describes the primary gene transcript and tentatively suggests the presence of alternative transcripts.

### *A. thaliana*: AT1G68945 – annotation and data inconsistent

Figure 5 shows *AT1G68945* which has been confirmed as protein coding from cDNA and EST data, although the protein product has yet to be characterized. It has only one annotated gene model, comprising a long 5' UTR, a single coding exon, and a short 3' UTR. No significant DRS peaks are found associated with this gene model or within the 300bp window downstream of the 3' end of the current annotation and so the algorithm described in [19] does not re-annotate this gene and leads to the conclusion that it is not expressed. Curiously however, a strong signal is seen in the DRS data on the opposite strand, at the start of the 5' UTR annotation. This peak is broad, covering ~20bp, suggesting multiple possible poly(A) sites. Reads from the un-stranded RNA-seq data align precisely to the gene position confirming its location but not which strand it is on. One possible interpretation of this region is that there is a gene on the reverse strand that is not annotated in TAIR10 (as suggested in Table 5) this is also supported by single-stranded RNA-Seq data from the Ecker Lab [35]. However, the reverse strand in this region of the current genome build contains multiple stop codons suggesting it is unlikely to represent a single protein coding gene.

## Section 2: Disentangling gene expression in complex genomic regions

### Homo sapiens: Mettl12

Figure 6 illustrates the genomic region around the gene *Mettl12* which is located on the forward strand of chromosome 11. This region shows the challenges of annotation and expression quantification in complex regions and how combining different datasets, in particular strand-specific data that defines 3'-ends, can help alleviate some of these difficulties.

Ensembl v69 provides several different gene annotation models for *Mettl12*, while RefSeq reports a single gene model that is significantly different to the Ensembl annotations. All these models agree on a 5' UTR that includes an intron, within which resides a copy of the snoRNA, *snorna57* (this is one of four copies of this snoRNA that occur in the human genome). The Mettl12 locus is additionally complicated by the presence of a large protein-coding ORF, *C11orf48*, on the antisense strand that overlaps *Mettl12* completely. Ensembl provides a total of thirteen different gene models for *C11orf48*, while RefSeq lists a single gene model. In addition, the annotated 5' UTRs of several *C11orf48* gene models overlap with the 5' UTR of the forward strand ORF *C11orf83*, which itself has two separate gene models. The details of all these annotations are provided in Table 6.

As one might anticipate for such a complex region, the un-stranded Illumina RNA-seq data for this region are ambiguous, so quantifying gene expression from these data is problematic. The terminal four exons of *C11orf48* are strongly-expressed (read depth ~150-300) suggesting that the gene model *ENST00000524958* is the predominant expressed form of *C11orf48* in these data. This is reinforced by reads that map across the intron/exon boundaries for this gene model. Importantly, there are no reads mapping across any splice junctions immediately prior to the start of this annotation, clearly delineating this model from the others for *C11orf48*. Similarly, two exons of *C11orf83* are also strongly-expressed and show a consistent splicing pattern, but the expression appears to be truncated at a position that is inconsistent with all the current 3' UTR annotations for *C11orf83*, suggesting a possible new gene model for this gene. The picture in the intervening region, which covers *Mettl12*, *snora57* and another gene model for *C11orf48*, is far less clear. The low-level expression in this region shows little in the way of distinct exon/intron boundaries that would help to identify the origin for this expression, but marginal evidence for some other transcripts of *C11orf48* and for expression from *Mettl12* can be identified from individual reads that map across appropriate exon/intron boundaries.

In contrast, the DRS data are more straightforward to interpret and quantify, since they reliably identify the sequenced strand. Hence, they can be used to help inform the gene annotations and quantify the gene expression in human skin within this genomic region. The DRS data have four distinct sites of expression; three on the forward strand (Figure 6, Track A, 1-3) and one on the reverse strand (Figure 6, Track K). On the forward strand, peaks 1 & 2 coincide with the *MettI12* annotations. Peak 1 is located in the 5' UTR of the annotations but downstream of *snorna57* suggesting that this peak represents expression of the snoRNA precursor rather than the gene. Peak 2 is located in the annotated 3' UTR of *MettI12*, however it is only 13 bp downstream of the stop codon. The sequence in this region does not show any strong candidates for internal priming and the upstream sequence contains a slight variation on the canonical poly(A) motif (ATTAAA) 17 bp upstream. Although this signal hints at a new gene model for *MettI12,* with a short 3' UTR, the low-level of the expression makes this inconclusive. Further downstream on the forward strand, *C11orf83* is strongly expressed in the DRS data (peak 3), again with an apparently shorter 3' UTR than annotated. The details of all these novel transcript annotations are provided in Table 6. The data are not as clear for the reverse strand. Assuming the current annotations are correct, the exquisite positional precision of the DRS data and the lack of any DRS peaks at other locations on the reverse strand, suggest four strong gene-model candidates. Of these, model *ENST00000524958* is consistent with the DRS data and the RNAseq data, supporting the conclusion that this is the predominantly expressed form of *C11orf48*, at least in these samples.  The other potential models may be correct, just not expressed in these samples.

## Homo sapiens: RPL31

Figure 7 illustrates another example of a complex genomic region with ambiguous expression for convergent genes on opposite strands. On the forward strand, *RPL31* has eleven gene models annotated in Ensembl, and three annotated in RefSeq. Across these models, the 5' UTR is annotated with seven different start positions and the 3' UTR is annotated with twelve alternative end positions. On the reverse strand, *TBC1D8* is similarly complex, with ten Ensembl gene models and one RefSeq, four of which overlap with the five longest forms of *RPL31*.

Again, as one might expect for such a complex region, the RNA-seq data are ambiguous. The RNA-seq data include five strong peaks that echo the exon and UTR structure of five of the RPL31 gene models, however the considerable low-level expression covering much of the region makes it hard to draw firm conclusions from RNA-seq data alone. This ambiguity is dramatically reduced with the addition of the DRS data. In the DRS data, a strong signal is observed coincident with the downstream edge of the fifth RNA-seq peak (Figure 7, Track A, peak 1). This broad peak covers ~20bp and encompasses the 3' UTR ends of seven of the annotated models. The sequence immediately upstream of the peak is strongly AT rich, suggesting that the location of the poly(A) site in *RPL31* may not be very precisely controlled. Instead, a range of possible poly(A) positions occur with different likelihoods within this window.

Interestingly, two small peaks also occur in the DRS data further downstream (Figure 7, Track A, peaks 2 & 3), close to three of the longer *RPL31* gene models. The first of these extends the nearest gene model by 56 bp, the second lies within 5 bp of the end of the longest annotated model. Both of these peaks have the AATAAG variant of the canonical polyadenylation signal ~19 bases upstream of the peak. This weak but distinct signal clearly demonstrates that the shorter *RPL31* gene models are not the only form of transcripts made from this gene in these data. On the reverse strand, the DRS data shows a strongly expressed peak (Peak 5) that is coincident with the end of the 3' UTR annotated in the RefSeq *TBC1D8* gene model. However, a second peak ~1.2kb further downstream (Peak 4), identifies a new putative polyadenylation site for this gene. Both of these peaks show the polyadenylation motif AATAAG ~20 bp upstream. Accordingly, a new gene model is proposed for RPL31 that results in a transcript that overlaps with all the *RPL31* gene models.

## Section 3: A clearer picture of small RNA expression

It is currently not possible to quantify the expression of long and short RNAs in a single RNA-Seq experiment. In order to identify expression of mature miRNAs, in particular, a protocol is used that specifically selects very short (<30bp) RNA species and so excludes the ~200 bp fragments commonly selected by RNA-seq protocols. Mature intergenic miRNAs are ~21bp single stranded RNA molecules processed out of pre-miRNA hairpin loops found in pri-miRNA transcripts and are transcribed by RNA polymerase II ([36]). The pri-miRNAs have been shown to be polyadenylated via a variety of methods including PCR primers ([36-38]), sequence analysis ([39]) and sequencing ([40]). The miRNA*, which is not loaded in the RISC complex is not normally retained, but can often be observed in high-throughput sequencing.

miR-200c and miR-141 illustrate the advantages of combining DRS and RNA-seq data with small RNA-seq (sRNA-seq) data for a better characterisation of intergenic pri-miRNAs. Figure 8 shows the genomic region around miR-200c and miR141. This region is flanked by genes that are expressed in the DRS and RNA-seq data; *PTPN6* on the forward strand (Figure 8, Track D) and *PHB2* on the reverse strand (Figure 8, Track H). Aligning directly with the miRNA annotations are two pairs of peaks in the sRNA-seq data (Figure 8, Track G, 1 & 2) that correspond to the mature miRNAs miR-200c-5p/3p and miR-141-5p/3p sequences. In each case, the 3p sequences are the dominant expressed form, as shown by the relative heights of the sRNA-seq peaks within each pair.

The structure and extent of the pri-miRNA is clearly delineated by the RNA-seq data (Figure 8, Track E) in the regions flanking the two mature miRNA loci. No reads are detected within the intronic region that covers the pre- or mature miRNA regions suggesting that the pre-miRNAs processing and cleavage occurs rapidly, leaving the 5' and (polyadenlyated) 3' end fragments to be slowly degraded. The DRS data support this picture showing a cluster of expression ~200 bp downstream of miR-141-3p on the forward strand (Figure 8, Track A, 3) that has the tandem polyadenylation site motif AATAAATAAA 26 bp upstream.

## Section 4: Novel gene discovery

In addition to improving existing annotations, the combination of DRS, RNA-seq and other datasets also identifies and characterises genomic regions containing new feature candidates. The discovery of potential new snoRNAs in the downstream region of the gene *AT4G10810* in *A. thaliana*, shown in Figure 9, is an example. The RNA-seq data downstream of *AT4G10810* shows significant low-level expression over a ~600 bp region, with no strong evidence for intron/exon structure (Figure 9, Track E, 1). The DRS data in this region are complex, showing a considerable number of small peaks that suggest multiple possible alternative polyadenylation sites (Figure 9, Track K, 2). Combined, these imply a cluster of short, currently un-annotated, features. This picture is reinforced by the large peak in expression seen in the sRNA-seq data in this region (Figure 9, Track G, 3). This peak does not show the two-peak structure characteristic of mature miRNA sequences (see Section 4), leaving us to speculate on the nature of this short feature. The *SnoSeeker* (v1.1, [41]) snoRNA prediction algorithm predicts a snoRNA coincident with this position suggesting that this is a previously undiscovered snoRNA gene.

# Discussion

Detailed, complete, genomic feature annotations are a cornerstone of modern biology. Their importance, particularly for experiments that rely on high-throughput transcriptomics, cannot be overstated. However, defining these annotations is not a trivial task and is made more difficult by the fact that there may be multiple 'correct' annotations for a gene. While the importance of accurate annotations is widely recognised, the impact that alternative individual annotation, or an alternative set of annotations, has on the subsequent downstream analysis (*e.g.*, differential gene expression) and biological understanding is less well appreciated. Two distinct classes of problem occur commonly for genome annotations; an incomplete set of feature annotations and/or an unreliable individual feature annotation.

The known set of human genes is an example of an incomplete set of feature annotations, *i.e.* a set of individual annotations (each of which may also be incomplete), that is missing discrete members of the set. Over the past decade considerable effort has been expended in manually curating the annotations for the human genome. As a consequence, the annotations for known genes is precise given the available data but the set as a whole is still likely to be missing as-yet undiscovered genes and alternatively processed mRNA isoforms ([29]). For human and other heavily curated genomes, even though the full set of information is not known, the information that exists for the individual annotations is often reliable. Providing the set is not too incomplete, it will have relatively little impact on downstream analyses that rely on these annotations. One important exception is where features that are not annotated overlap completely with known features. For example, the observed foldchange for such a region could be completely misleading and would not reflect the underlying biology if expression of the overlapping genes is very different.

Unreliable individual annotations present a different challenge. Here, members of a set of feature annotations (that may be partially complete) are based on a limited or significantly imprecise set of information. The impact this has on any downstream data analyses depends on the properties of the data being used and the specific analyses. For example, differential gene expression between two experimental conditions based on RNA-seq data is not dramatically sensitive to having marginally inaccurate annotation of gene structure unless the gene structure changes between conditions. Since the conditions being compared both use the same annotations, and given that the annotations are covered by a significant majority of the reads, the calculated fold change will be similar to the actual fold change that would be calculated using a more accurate set of annotations. Techniques that focus on one region of the gene such as DRS are far more sensitive to inaccurate or incomplete annotation information. If the locus that has been sequenced is not included within the annotated GARs of the gene then no (or very little) expression will be attributed to this gene in either condition, regardless of the true change in expression in the data.

For most published genomes, the available annotation is the result of an automated prediction-based annotation pipelines (see, for example, [42], [43]). Automated gene prediction is a difficult challenge (see [44]) and these first-pass annotations often contain considerable inaccuracies. Re-annotation using automatic methods typically involved discarding the current set of annotations and building the annotations again from scratch as the genome sequence is improved. In some cases, re-annotation has been attempted by supplementing the current annotations guided by high-throughput transcriptomics sequencing data ([19]). Automated, but data-driven, re-annotations can provide a considerable increase in the quality of feature annotations however they still have several drawbacks. Typically automatic methods depend on several arbitrarily set parameters such as the size of the window probed for new feature endpoints and the minimum number of reads required to extend an annotation (this is also true of automated annotation pipelines). As a result, many individual feature annotations will remain inaccurate and/or the annotation set remain incomplete. The *A. thaliana* re-annotation provided by [19] considerably extends and improves on an already comprehensive and detailed genome annotation in a well-studied model species (TAIR version 10 - [45]). However, the automated annotation method is unable successfully to re-annotate genes requiring a 3' extension longer than the 300 bp downstream window, nor can it distinguish between a genuine new 3' end annotation or the 3' end of a new short gene located immediately downstream of an existing annotation (see, for example, Section 5 and Figure 9). Even after re-annotation dozens of intergenic DRS peaks (many comprised of >50 raw reads) remain un-accounted for, indicating the need for a more careful data-driven re-annotation.

The majority of high-throughput transcriptomics sequencing datasets are not generated with the primary intention of re-annotating genomic features, yet these datasets provide a wealth of information that can do exactly that. Individual sequencing technologies often show characteristics that make it difficult to base strong conclusions about feature re-annotation solely on the data they generate (Table 9). The experience gained in the present study suggest that genome annotation efforts that focus on using a single data type (for example, [46]) are likely to have difficulty producing a high-quality, high-completeness set of feature annotations for eukaryotic genomes. Combining the strengths of RNA-seq data, short RNA-seq, archival EST/mRNA data and strand-specific sequencing that defines the 3'-end is particularly effective at overcoming the weaknesses inherent to data generated from any one of these technologies individually (Table 5). These data can be used to identify and characterise gene intron/exon structure, and characterise GARs associated with these genes. The DRS data is particularly important in this context, both by providing precise information about the termination point of 3' UTRs and by unambiguously identifying the strand for the gene expression data. Accurately constraining 3' UTRs associated with genes is particularly important for alternative polyadenylation studies, microRNA and other regulatory element binding site identification. It is also important for downstream differential gene expression analysis and functional pathway analysis, because a significant fraction of RNA-seq reads, and all DRS reads, associated with a gene lie within their associated 3'UTR.

Careful re-annotation of genome features from data such as these holds great potential for novel discoveries in addition to improving the quality and reliability of every scientific result which builds on the re-annotated features. The examples presented here are entirely data-driven, removing the need to rely on computational predictions. However, this re-annotation process is not always straightforward even with complementary data sets and it has proven to be difficult to automate effectively (particularly compared to standard gene prediction routines). It is clear that automatic annotation pipelines will improve with the inclusion of strand-specific RNA-seq data and data that delineates the 5' and 3' ends precisely. Indeed, major projects such as Ensembl are now incorporating these data into their annotation pipelines (S. Searle per. Comm.). However, the examples presented in this paper suggest that for complete and precise annotation there is currently no substitute for annotation curated by experienced and knowledgeable scientists from a combination of DRS, RNA-seq, sRNA-seq, EST and other informative data.

## Figure Legends

**Figure 1. The genomic context around *BMPR1A* in *G. gallus*.**

Figures 1-8 are divided into three regions comprising information located on the forward strand (pink), reverse strand (grey) and un-stranded information (yellow). Each region is subdivided into tracks showing a selection of the different annotations/datasets described below. For clarity, tracks are omitted where the track contains no data in the region shown.

**Tracks A & K:** Histograms for forward (A) and reverse (K) strands computed by summing the number of uniquely aligned DRS reads that end at a position and presented in units of read-counts/base.

**Tracks B & J:** Filled rectangles show forward (B) and reverse (J) strand individual EST alignments for a selection of the total EST coverage. Individual EST alignments that span across an implied exon splice junction are illustrated by a split bar representing the sequenced EST joined by a thin line that spans the implied intron.

**Tracks C & I:** Additional annotation information for forward (C) and reverse (I) strands. This track shows annotation information that doesn't originate from a primary reference database for the species. Details of the specific annotations shown for each figure are given in the figure caption.

**Tracks D & H:** Primary database annotations labelled with the database primary identifier for forward (D) and reverse (H) strands. Multiple gene models are shown where appropriate. Exons are shown as thick bars, UTRs as thinner bars and introns as thin lines. For *A. thaliana* this track shows the TAIR (v10) annotations. For the other examples in this paper, this track shows Ensembl (v69, red) and RefSeq (v191, green) annotations.

**Track E:** Unstranded RNA-seq read depth histogram, computed by summing the number of uniquely aligned reads that cover at any given position and expressed in read counts/base.

**Track F:** RNA-seq individual read alignments, for a selection of the total read depth, shown as filled rectangles. Individual read alignments that span across an implied exon splice junction are represented by a split bar representing the sequenced read joined by a thin line showing the implied intron.

**Track G:** Unstranded sRNA-seq read depth histogram, computed by summing the number of uniquely aligned sRNA-seq reads that cover at any given position and expressed in units of read-counts/base.

Figure 1 shows a ~57kb region of *G. gallus*, chromosome 6, including *BMPR1A* (*ENSGALT00000003119*) and illustrates a straight-forward gene re-annotation, where the RNA-Seq and DRS data combined are sufficient to define the extent, structure, and alternative polyadenylation positions for a gene. Tracks C & I show confirmed complete coding sequence mRNA data for the region (GenBank v191 - orange) and the locations of the Affymetrix chicken GeneChip microarray probe-sets (black markers), and the cDNA against which the Affymetrix probe-sets were designed (light blue). See Supplementary Data Sections 1 and 2 for more details on the generation and processing of the RNA-seq and DRS data-sets. The EST data (B & J) are from [47]. The DRS track for the reverse strand (Track H) contains no data in the region shown and has been removed for clarity.

**Figure 2. The genomic context around *HOXA7* in *G. gallus*.**

The individual tracks and layout of this figure are as described in Figure 1. Figure 2 shows a ~6kb region of *G. gallus*, chromosome 2 that encompasses *HOXA7* gene. The RNA-seq (Tracks E & F), Helicos BioSciences' DRS (Tracks A & K) and publically available EST (Tracks B & J) datasets for this region are ambiguous, but combined, the data clearly define the extent, and structure for this gene. Tracks C & I show the same additional annotation tracks as shown in Figure 1. See Supplementary Data Sections 1 & 2 for more details on the generation and processing of the RNA-Seq and DRS data-sets. EST data were taken from [47].

**Figure 3. The genomic context around *SLFN5* in *H. sapiens*.**

This figure shows a ~6kb region of *H.sapiens*, chromosome 17, that encompasses the recently re-annotated *SLFN5* gene. Two peaks in the DRS data for this region (Track A) reveal that even our most up-to-date annotations in heavily curated genomes are often incomplete. The difference between the annotations provided by RefSeq and Ensembl (Track D) also highlights that existing primary database annotations often disagree significantly, making downstream analysis results dependent of the reference database used for individual studies. For full details of the individual tracks and layout of this figure, see the legend to Figure 1. See Supplementary Data Sections 3 & 4 for more details on the generation and processing of the RNA-seq and DRS data-sets.

**Figure 4. The genomic context around *AT4G02715* in *A. thaliana*.**

A ~3kb region of A. thaliana on chromosome 4 is shown, which encompasses AT4G02715. In this case the extensive 3' UTR extension suggested by the DRS data (Track K) shows how this re-annotation was missed even by the automated re-annotation algorithm applied in [19]. For full details of the individual tracks and layout of this figure, see Legend to Figure 1. See Supplementary Data Sections 6, 7 & 8 for more details on the RNA-seq, EST and DRS data-sets, and their processing.

**Figure 5. The genomic context around *AT1G68945* in *A. thaliana*.**

This figure shows a ~600 bp region of *A. thaliana*, chromosome 1, around the existing annotation of the gene *AT1G68945*. In this case, the DRS data for this region (Track K) reveal that the existing annotation is on the incorrect strand. This kind of situation is difficult for automated re-annotation pipelines to deal with, particularly if they focus on using natively un-stranded data, such as Illumina RNA-Seq, to inform the annotation. This highlights necessity of natively stranded data, such as DRS data, for correctly defining feature annotations. For full details of the individual tracks and layout of this figure, see Figure 1 (caption). See Supplementary Data Sections 6, 7 & 8 for more details on the RNA-Seq, EST and DRS datasets, and their processing.

**Figure 6. The genomic context around *Mettl12* in *H. sapiens.***

This figure shows a complex region of the human genome that is difficult to annotate either automatically or manually. The combination of DRS and RNA-Seq data for this ~13kb region of *H. sapiens*, chromosome 11, brings greater clarity to the feature annotation in this region, that either dataset individually is incapable of providing. In particular, the DRS data on the forward strand (Track A) clearly identifies the expression of *snoRNA57*, in the first intron of *Mettl12*, and several new transcripts for both *Mettl12* and *C11orf83*. The combination of the exon structure seen in the RNA-seq data (Tracks E & F) and the DRS datal on the reverse strand (Track K) clearly identify the dominant form of *C11orf48* observed in these data. For full details of the individual tracks and layout of this figure, see Figure 1 (caption). See Supplementary Data Sections 3 & 4 for more details on the RNA-Seq and DRS data-sets, and their processing.

**Figure 7. The genomic context around *RPL31* in *H. sapiens*.**

This ~25kb region of *H. sapiens*, chromosome 2, again highlights the difficulties in interpreting unstranded data in complex genomes. This region encompasses the gene *RPL31* on the forward strand and *TBC1D8* on the reverse strand. Many of the existing annotations for these two genes overlap (Tracks D & H) making unstranded data difficult to interpret with certainty. The natively stranded DRS data (Tracks A & K) clearly delineate the ends of the transcripts observed from both these genes, including a new annotation for *TBC1D8*.  For full details of the individual tracks and layout of this figure, see Figure 1 (caption)**.** See Supplementary Data Sections 3 & 4 for more details on the RNA-seq and DRS data-sets, and their processing.

**Figure 8. The genomic context around hsa-mir-200c~141 in *H. sapiens*.**

It is currently not possible to quantify the expression of both long and short RNAs in a single RNA-seq experiment making it difficult to get a complete picture of miRNA transcription. In this example, the combination of DRS (Track A), RNA-seq (Tracks E & F) and sRNA-seq (Track G) datasets shows the extent of the pri-miRNA that codes for miR-200c and miR-141. The lack of reads detected in the intronic region of the pri-mRNA in the RNA-seq data also suggests that the pri- and pre-miRNA processing stages occur rapidly. See Supplementary Data Sections 3, 4 and 5 for more details on the RNA-seq, DRS and sRNA-seq data-sets, and their processing.

**Figure 9. The genomic context around *AT4G10810* in *A. thaliana*.**

This figure shows a ~2kb region of *A. thaliana,* chromosome 4, including *AT4G10810* that demonstrates the capability of combined DRS, RNA-seq and sRNA-seq to identify novel genes. This also highlights some of the limitations of automated re-annotation algorithms that are based on arbitrarily chosen parameter values. In this case, [19] (2012), provide a re-annotation of the 3' UTR of *AT4G10810* by focussing on the DRS data within a region 300bp downstream of the end of the primary database annotations (Track K). For most *A. thaliana* genes, this proves to be an effective strategy, but occasionally it results in incorrect re-annotations. Here, the region downstream of *AT4G10810* encompasses multiple relatively weak DRS peaks (Track K, 2) and Sherstnev *et al* mistakenly re-annotate the gene to include many of these peaks (Track I). In fact, the RNA-seq data (Tracks E & F, 1) clearly identify the spatial separation between *AT4G10810* and the significant low-level downstream expression, suggesting a novel gene, or cluster of genes. Interestingly, a strong peak in the sRNA-seq data in this region (Track G, 3), coupled with a coincident prediction from SnoSeeker (Track I), strongly suggests the presence of a novel snoRNA in this region. See Supplementary Data Sections 6, 7 & 8 for more details on the generation and processing of the RNA-seq, sRNA-seq, EST and DRS data-sets.

## Tables

**Table 1. Comparison of annotations for *BMPR1A*.**

| Primary annotation | Chr | Begin (bp) | End (bp) | Strand | Coverage (bp) |
|---|---|---|---|---|---|
| RefSeq: *BMPR1A* | 6 | 3,546,262 | 3,585,602 | + | 39,340 |
| ensembl: *ENSGALT00000003119* | 6 | 3,546,283 | 3,585,813 | + | 39,530 |
| **Proposed re-annotation** | | | | | |
| EST/RNA-seq: 5' UTR | 6 | 3,546,262 | 3,564,179 | + | 17,917 |
| EST/RNA-seq: *BMPR1A* | 6 | 3,564,180 | 3,585,585 | + | 21,405 |
| DRS/EST/RNA-seq: 3' UTR | 6 | 3,585,586 | 3,590,064 | + | 4,478 |
| Summary | 6 | 3,546,262 | 3,590,064 | + | 43,800 |

**Table 2. Comparison of annotations for *HOXA7*.**

| Primary annotation | Chr | Start (bp) | Stop (bp) | Strand | Coverage (bp) |
|---|---|---|---|---|---|
| RefSeq: *HOXA7* | 2 | 32,570,322 | 32,572,159 | - | 1,837 |
| ensembl: *ENSGALT00000018013* | 2 | 32,570,285 | 32,571,987 | - | 1,702 |
| **Proposed re-annotation** | | | | | |
| EST/RNA-seq: 5' UTR | 2 | 32,572,160 | 32,572,292 | - | 132 |
| EST/RNA-seq: *HOX7A* | 2 | 32,570,322 | 32,572,159 | - | 1,837 |
| DRS/EST/RNA-seq: 3' UTR | 2 | 32,568,768 | 32,570,321 | - | 1,553 |
| Summary | 2 | 32,572,160 | 32,570,321 | - | 3,522 |

**Table 3. Comparison of annotations for *SLFN5* gene locus.**

| Primary annotation | Chr | Start (bp) | End (bp) | Strand | Coverage (bp) |
|---|---|---|---|---|---|
| RefSeq: *SLFN5* (NM_144975) | 17 | 33,570,086 | 33,594,768 | + | 24,682 |
| ensembl: *ENST00000299977* | 17 | 33,570,055 | 33,600,674 | + | 30,619 |
| ensembl: *ENST00000542451* | 17 | 33,570,090 | 33,593,379 | + | 23,289 |
| ensembl: *ENST00000299977* | 17 | 33,570,108 | 33,586,839 | + | 16,731 |
| **Proposed re-annotation 1** | | | | | |
| RNA-seq: 5' UTR | 17 | 33,570,055 | 33,585,708 | + | 15,653 |
| RNA-seq: *SLFN5* | 17 | 33,585,709 | 33,592,121 | + | 6,412 |
| RNA-seq/DRS: 3' UTR | 17 | 33,592,121 | 33,597,113 | + | 4,992 |
| Summary | 17 | 33,570,055 | 33,597,113 | + | 27,057 |
| **Proposed re-annotation 2** | | | | | |
| RNA-seq: 5' UTR | 17 | 33,570,055 | 33,585,708 | + | 15,653 |
| RNA-seq: *SLFN5* | 17 | 33,585,709 | 33,592,121 | + | 6,412 |
| RNA-seq/DRS: 3' UTR | 17 | 33,592,121 | 33,600,669 | + | 8,548 |
| Summary | 17 | 33,570,055 | 33,600,669 | + | 30,613 |

**Table 4. Comparison of annotations for *AT4G02715* gene locus.**

| Primary annotation | Chr | Start (bp) | End (bp) | Strand | Coverage (bp) |
|---|---|---|---|---|---|
| TAIR10: *AT4G02715* | 4 | 1,203,279 | 1,202,169 | - | 1,110 |
| **Proposed re-annotation 1** | | | | | |
| RNA-seq/EST: 5' UTR | 4 | 1,203,279 | 1,202,998 | - | 281 |
| RNA-seq/EST: *AT4G02715* | 4 | 1,202,998 | 1,202,169 | - | 829 |
| RNA-seq/DRS/EST: 3' UTR | 4 | 1,202,169 | 1,200,886 - 1,200,975 | - | 1194 - 1,279 |
| Summary | 4 | 1,203,279 | 1,200,886 - 1,200,975 | - | 2,304 - 2,389 |
| **Proposed re-annotation 2** | | | | | |
| RNA-seq/EST: 5' UTR | 4 | 1,203,279 | 1,202,998 | - | 281 |
| RNA-seq/EST: *AT4G02715* | 4 | 1,202,998 | 1,202,169 | - | 829 |
| RNA-seq/DRS/EST: 3' UTR | 4 | 1,202,169 | 1,200,688 | - | 1,481 |
| Summary | 4 | 1,203,279 | 1,200,688 | - | 2,591 |
| **Proposed re-annotation 3** | | | | | |
| RNA-seq/EST: 5' UTR | 4 | 1,203,279 | 1,202,998 | - | 281 |
| RNA-seq/EST: *AT4G02715* | 4 | 1,202,998 | 1,202,169 | - | 829 |
| RNA-seq/DRS/EST: 3' UTR | 4 | 1,202,169 | 1,200,666 | - | 1,503 |
| Summary | 4 | 1,203,279 | 1,200,666 | - | 2,613 |

**Table 5. Comparison of annotations for *AT1G68945* gene locus.**

| Primary annotation | Chr | Start (bp) | End (bp) | Strand | Coverage (bp) |
|---|---|---|---|---|---|
| TAIR10: *AT1G68945* | 1 | 25,926,962 | 25,927,330 | + | 368 |
| **Proposed re-annotation** | | | | | |
| RNA-seq/EST: 5' UTR | 1 | 25,927,329 | 25,927,314 | - | 15 |
| RNA-seq/EST: *AT1G68945* | 1 | 25,927,313 | 25,927,167 | - | 146 |
| RNA-seq/DRS/EST: 3' UTR | 1 | 25,927,166 | 25,926,947 - 25,926,967 | - | 199-219 |
| Summary | 1 | 25,927,329 | 25,926,947 - 25,926,967 | - | 360-380 |

**Table 6. Comparison of annotations for *Mettl12* gene locus.**

| Primary annotation | Chr | Start (bp) | End (bp) | Strand | Coverage (bp) |
|---|---|---|---|---|---|
| RefSeq: *Mettl12* | 11 | 62,432,779 | 62,434,923 | + | 2,145 |
| ensembl: *ENST00000532971* | 11 | 62,432,781 | 62,435,580 | + | 2,800 |
| ensembl: *ENST00000398922* | 11 | 62,432,781 | 62,434,869 | + | 2,089 |
| ensembl: *ENST00000529868* | 11 | 62,432,785 | 62,435,968 | + | 3,184 |
| **Proposed re-annotation** | | | | | |
| RNA-seq: 5' UTR | 11 | 62,432,794 | 62,433,350 | + | 557 |
| RNA-seq: *Mettl12* | 11 | 62,433,351 | 62,434,522 | + | 1,172 |
| RNA-seq/DRS: 3' UTR 1 | 11 | 62,433,867 | 62,434,535 | + | 4,992 |
| **Primary annotation** | | | | | |
| RefSeq: *snoRNA57* | 11 | 62,432,893 | 62,433,041 | + | 148 |
| Ensembl: *ENST00000206597* | 11 | 62,432,893 | 62,433,041 | + | 149 |
| **Additional annotation** | 11 | | | | |
| *snoRNA57* precursor | 11 | 62,432,794 | 62,433,179 | + | 385 |
| **Primary annotation** | | | | | |
| RefSeq: *C11orf83* | 11 | 62,439,125 | 62,441,161 | + | 2,036 |
| ensembl: *ENST00000531323* | 11 | 62,437,745 | 62,441,049 | + | 3,304 |
| ensembl: *ENST00000377953* | 11 | 62,439,126 | 62,441,159 | + | 2,033 |
| **Proposed re-annotation** | | | | | |
| RNA-seq: 5' UTR | 11 | 62,439,125 | 62,439,216 | + | 91 |
| RNA-seq: *C11orf83* | 11 | 62,439,217 | 62,439,584 | + | 367 |
| RNA-seq/DRS: 3' UTR 1 | 11 | 62,439,585 | 62,439,844 | + | 259 |
| Summary | 11 | 62,439,125 | 62,439,844 | + | 719 |

## Table 7. Transcript annotations for *RPL31* gene locus

| Primary annotation | Chr | Start (bp) | End (bp) | Strand | Coverage (bp) |
|---|---|---|---|---|---|
| RefSeq: *RPL31 NM_001098577.2* | 2 | 101,618,690 | 101,636,154 | + | 17,464 |
| RefSeq: *RPL31 NM_001099693.1* | 2 | 101,618,690 | 101,622,884 | + | 4,194 |
| RefSeq: *RPL31 NM_000993.4* | 2 | 101,618,690 | 101,622,884 | + | 4,194 |
| ensembl: *ENST00000264258* | 2 | 101,618,177 | 101,623,729 | + | 5,612 |
| ensembl: *ENST00000409320* | 2 | 101,618,755 | 101,622,880 | + | 4,125 |
| ensembl: *ENST00000409711* | 2 | 101,619,153 | 101,622,829 | + | 3,676 |
| ensembl: *ENST00000456292* | 2 | 101,619,153 | 101,622,533 | + | 3,380 |
| ensembl: *ENST00000409000* | 2 | 101,618,691 | 101,621,066 | + | 2,375 |
| ensembl: *ENST00000409028* | 2 | 101,618,745 | 101,636,078 | + | 17,333 |
| ensembl: *ENST00000409650* | 2 | 101,618,755 | 101,634,751 | + | 15,996 |
| ensembl: *ENST00000409038* | 2 | 101,618,755 | 101,634,768 | + | 16,013 |
| ensembl: *ENST00000409733* | 2 | 101,618,755 | 101,622,881 | + | 4,126 |
| ensembl: *ENST00000441435* | 2 | 101,619,201 | 101,640,494 | + | 21,293 |
| ensembl: *ENST00000419276* | 2 | 101,618,773 | 101,622,885 | + | 4,152 |
| **Proposed re-annotation 1** | | | | | |
| RNA-seq: 5' UTR | 2 | 101,618,690 | 101,619,162 | + | 472 |
| RNA-seq: *RPL31* | 2 | 101,619,163 | 101,622,842 | + | 3,679 |
| RNA-seq/DRS: 3' UTR 1 | 2 | 101,622,843 | 101,622,865 - 101,622,887 | + | 22 - 44 |
| Summary | 2 | 101,618,690 | 101,622,865 - 101,622,887 | + | 4,175 - 4,197 |
| **Proposed re-annotation 2** | | | | | |
| RNA-seq: 5' UTR | 2 | 101,618,690 | 101,619,162 | + | 472 |
| RNA-seq: *RPL31* | 2 | 101,619,163 | 101,635,499 | + | 3,679 |
| RNA-seq/DRS: 3' UTR 1 | 2 | 101,635,500 | 101,636,201 | + | 22 - 44 |
| Summary | 2 | 101,618,690 | 101,636,201 | + | 4,175 - 4,197 |
| **Proposed re-annotation 3** | | | | | |
| RNA-seq: 5' UTR | - | - | - | - | - |
| RNA-seq: *RPL31* | 2 | 101,619,201 | 101,640,097 | + | 20,896 |
| RNA-seq/DRS: 3' UTR 1 | 2 | 101,640,098 | 101,640,488 | + | 390 |
| Summary | 2 | 101,619,201 | 101,640,488 | + | 21,287 |
| **Primary annotation** | | | | | |
| RefSeq: *TBC1D8* | 2 | 101,623,690 | 101,767,846 | - | 4,163 |
| Ensembl: *ENST00000409318* | 2 | 101,624,079 | 101,767,846 | - | 3,803 |
| **Proposed re-annotation** | | | | | |
| RNA-seq: 3' UTR | 2 | 101,622,395 | 101,624,281 | - | 1,886 |
| RNA-seq: *TBC1D8* | 2 | 101,624,282 | 101,767,714 | - | 143,432 |
| RNA-seq/DRS: 5' UTR | 2 | 101,767,715 | 101,767,730 | - | 15 |

**Table 8. Transcript annotations for *AT4G10810* gene locus**

| Primary annotation | Chr | Start (bp) | End (bp) | Strand | Coverage (bp) |
|---|---|---|---|---|---|
| TAIR10: *AT4G10810* | 4 | 6,646,335 | 6,645,715 | - | 620 |
| [19] | 4 | 6,646,335 | 6,645,421 | - | 914 |
| **Proposed re-annotation 1** | | | | | |
| RNA-seq/EST: 5' UTR | 4 | 6,646,335 | 6,646,229 | - | 106 |
| RNA-seq/EST: *AT1G68945* | 4 | 6,646,230 | 6,645,984 | - | 246 |
| RNA-seq/DRS/EST: 3' UTR | 4 | 6,645,985 | 6,645,715 - 6,645,864 | - | 121-270 |
| Summary | 4 | 6,646,335 | 6,645,715 - 6,645,864 | - | 471-620 |
| **Proposed re-annotation 2** | | | | | |
| Novel snoRNA | 4 | 6,645,422 | 6,645,529 | - | 107 |
| *snoSeeker* Predicted snoRNA | 4 | 6,645,420 | 6,645,538 | - | 118 |

**Table 9. Strengths and weaknesses of different data types.**

| | Strengths | Weaknesses |
|---|---|---|
| **EST** | Ubiquitous | Strande-ness is unreliable |
| | Confirmed transcripts | Low coverage |
| | Reveals gene structure | Biased sampling of the transcriptome |
| | Stranded | |
| **DRS** | Natively stranded | |
| | Exquisite positional accuracy (±2bp) | Only probes 3' UTR end position |
| | Quantitative | Cannot reveal gene structure |
| | No amplification | Short reads |
| | Unbiased sampling of the transcriptome | |
| **RNA-seq** | Easy/cheap to generate high coverage | Unstranded |
| | High sensitivity | Size selected (>200 bp) |
| | Reveals gene structure | Amplification step |
| | Unbiased sampling of the transcriptome | Sensitive to read alignment details |
| | Quantitative | |
| **sRNA-seq** | Sensitive only to small transcripts/exons | Unstranded |
| | Easy/cheap to generate high coverage | Size selected (~20 bp) |
| | High sensitivity | Amplification step |
| | Reveals miRNA structure | Sensitive to read alignment details |
| | Quantitative | No splicing |

**Bibiography**

Figure 1.

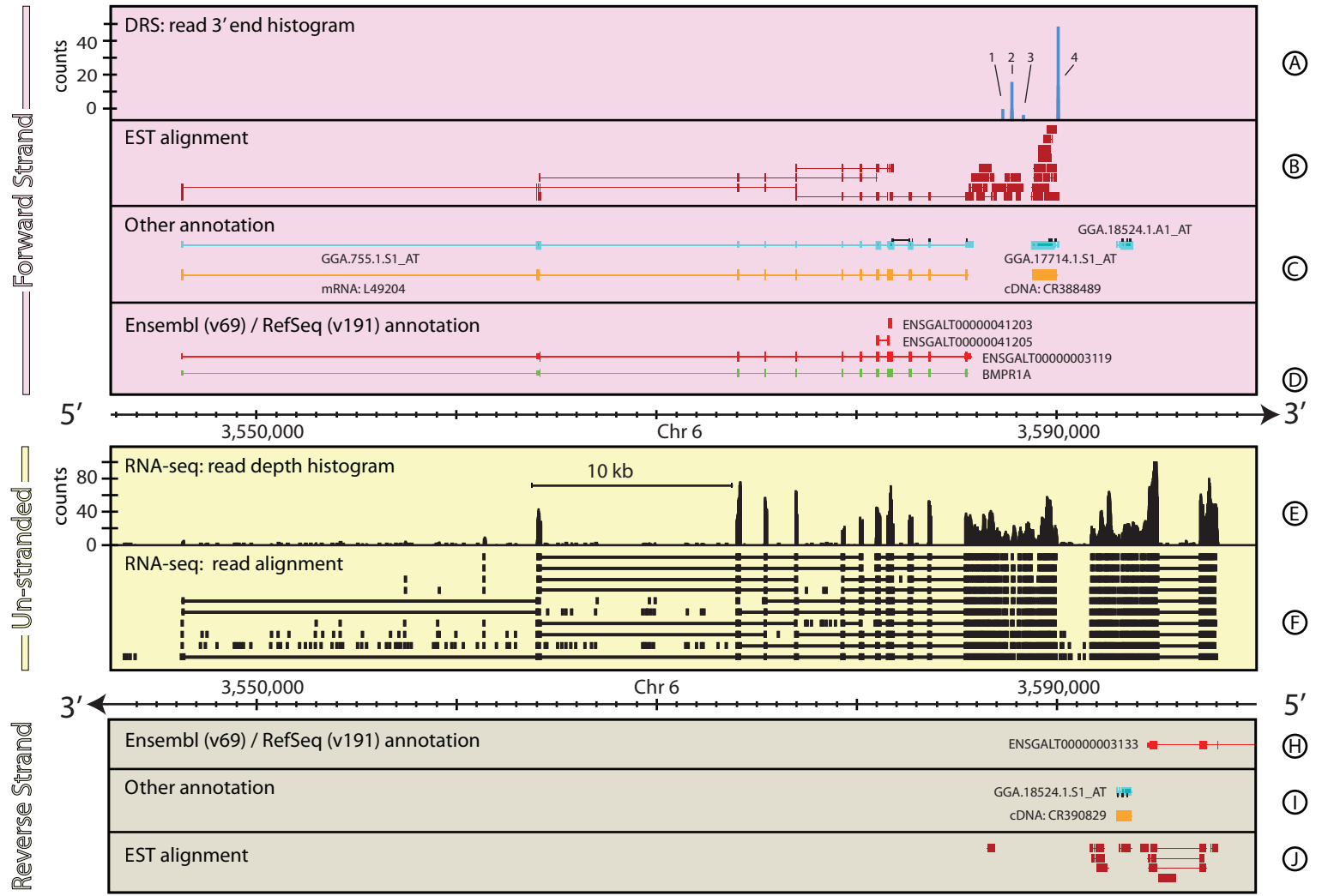

Figure 2.

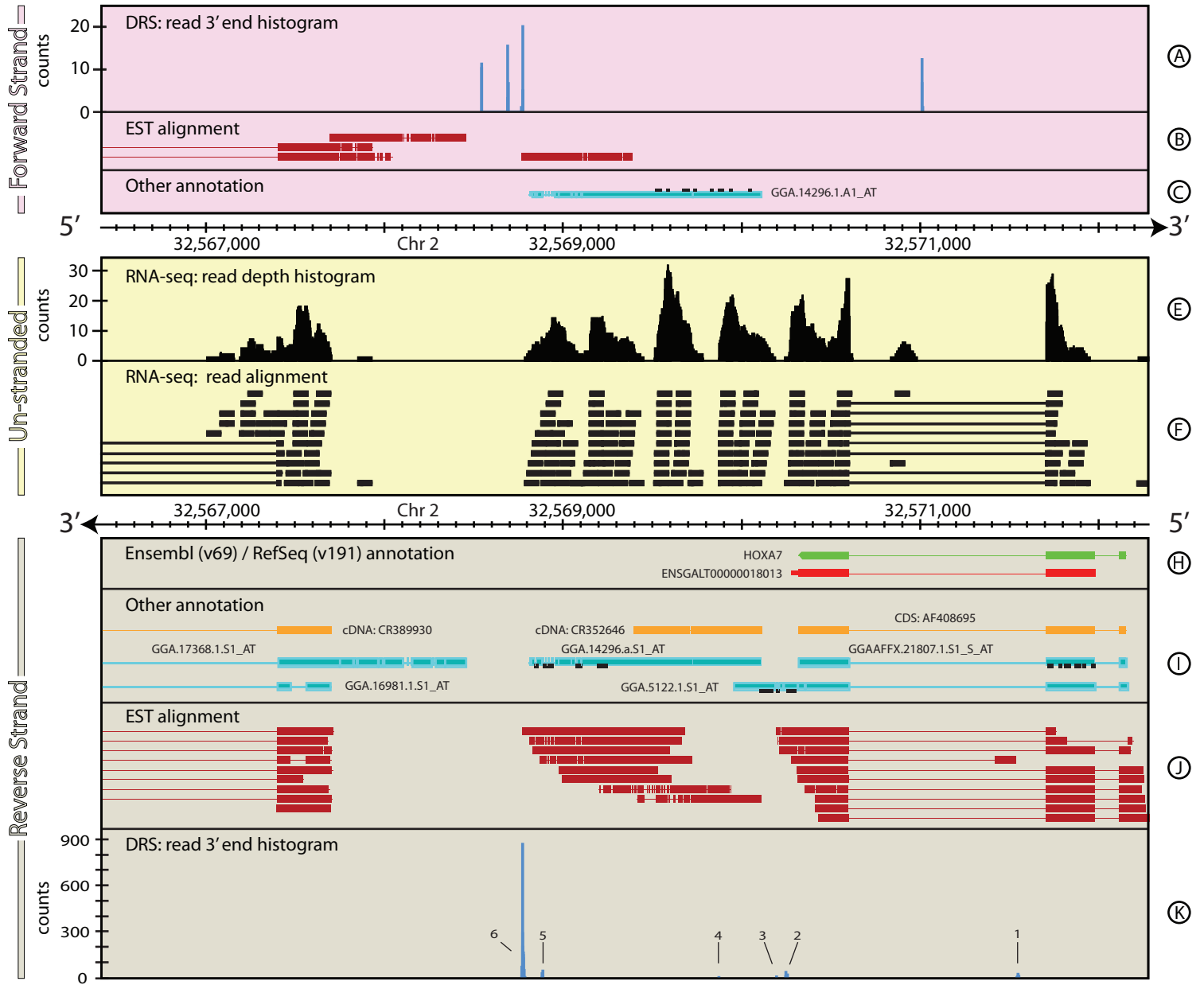

**Figure 3.**

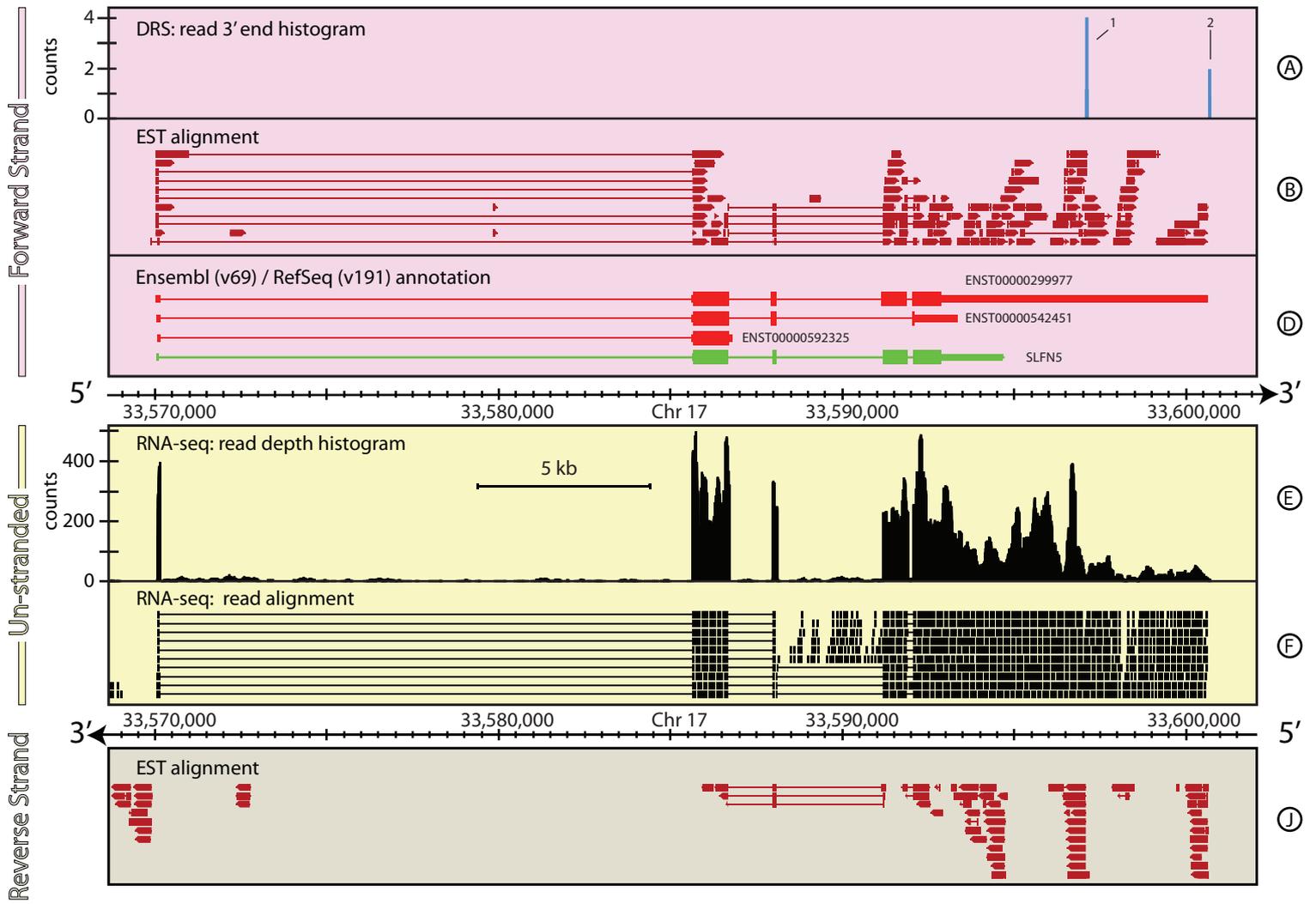

**Figure 4.**

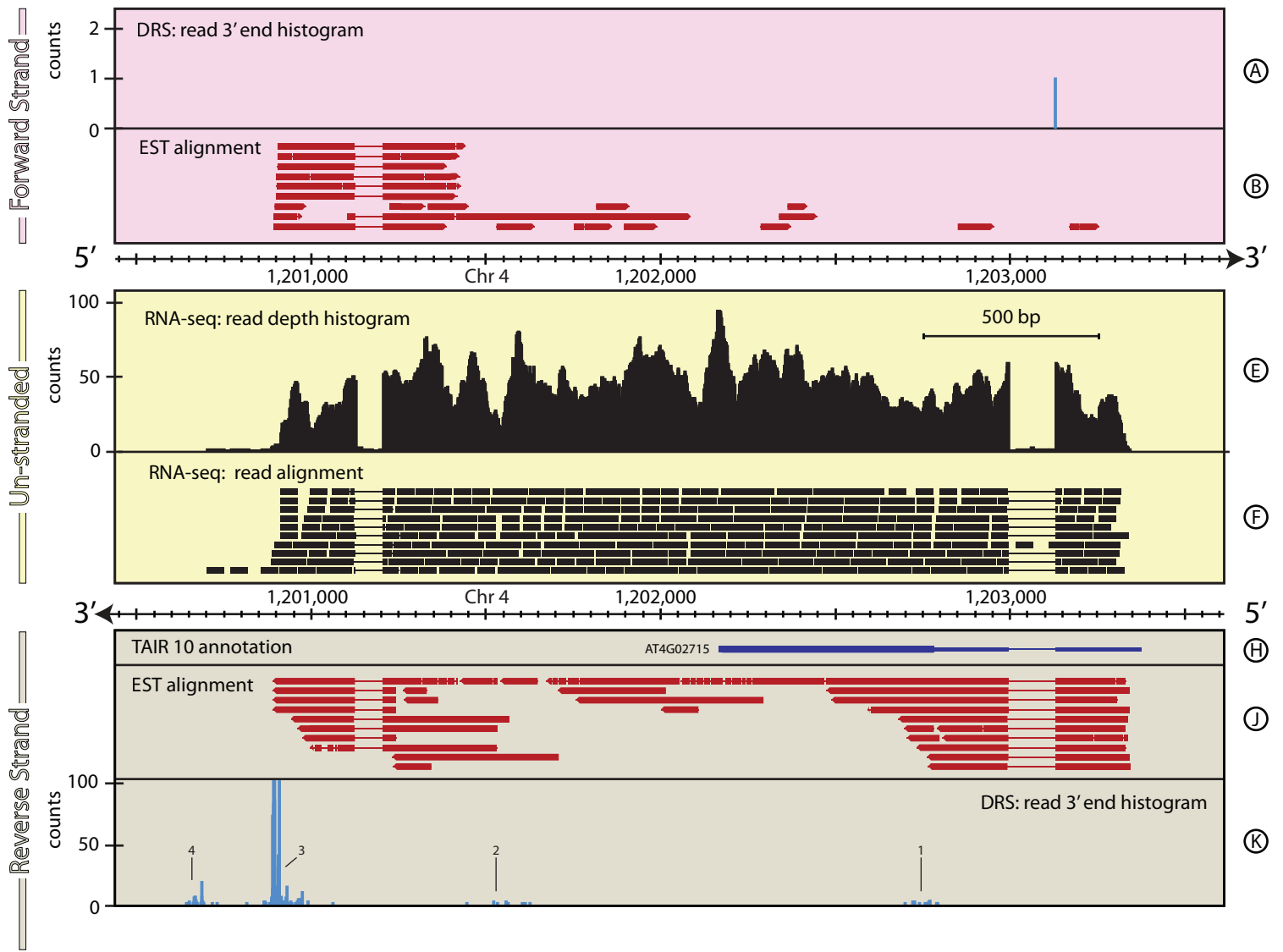

Figure 5.

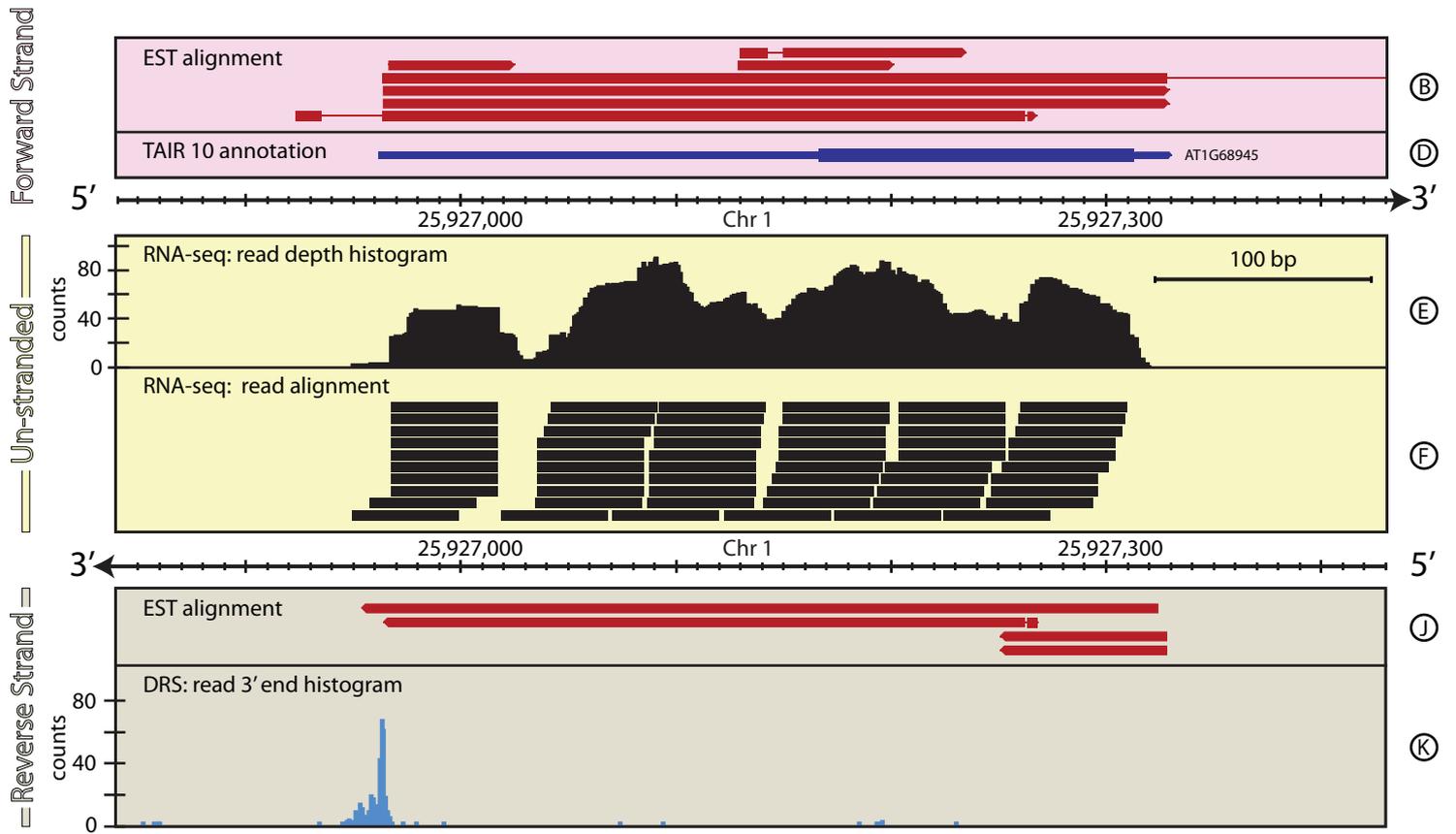

Figure 6.

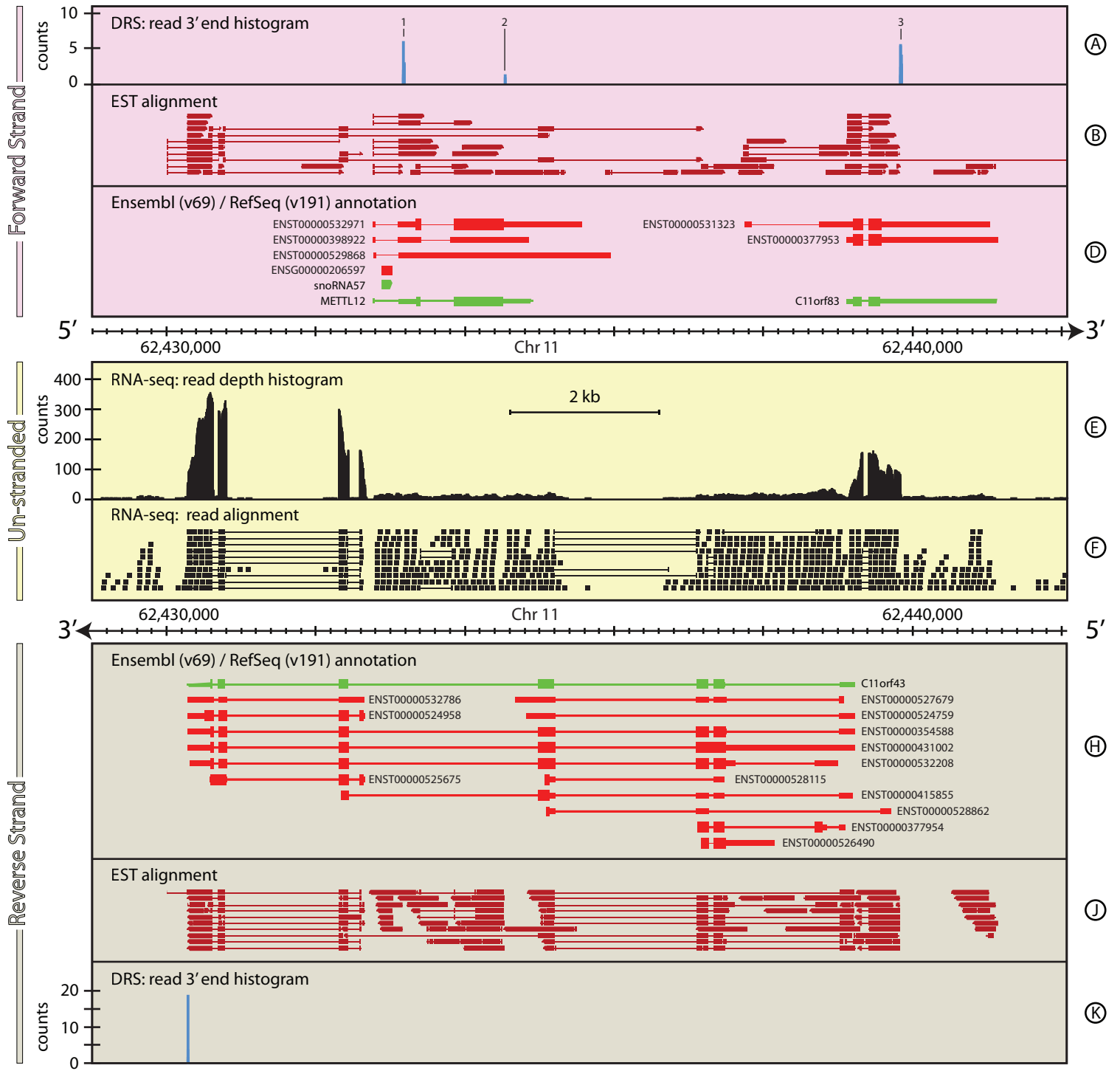

# Figure 7.

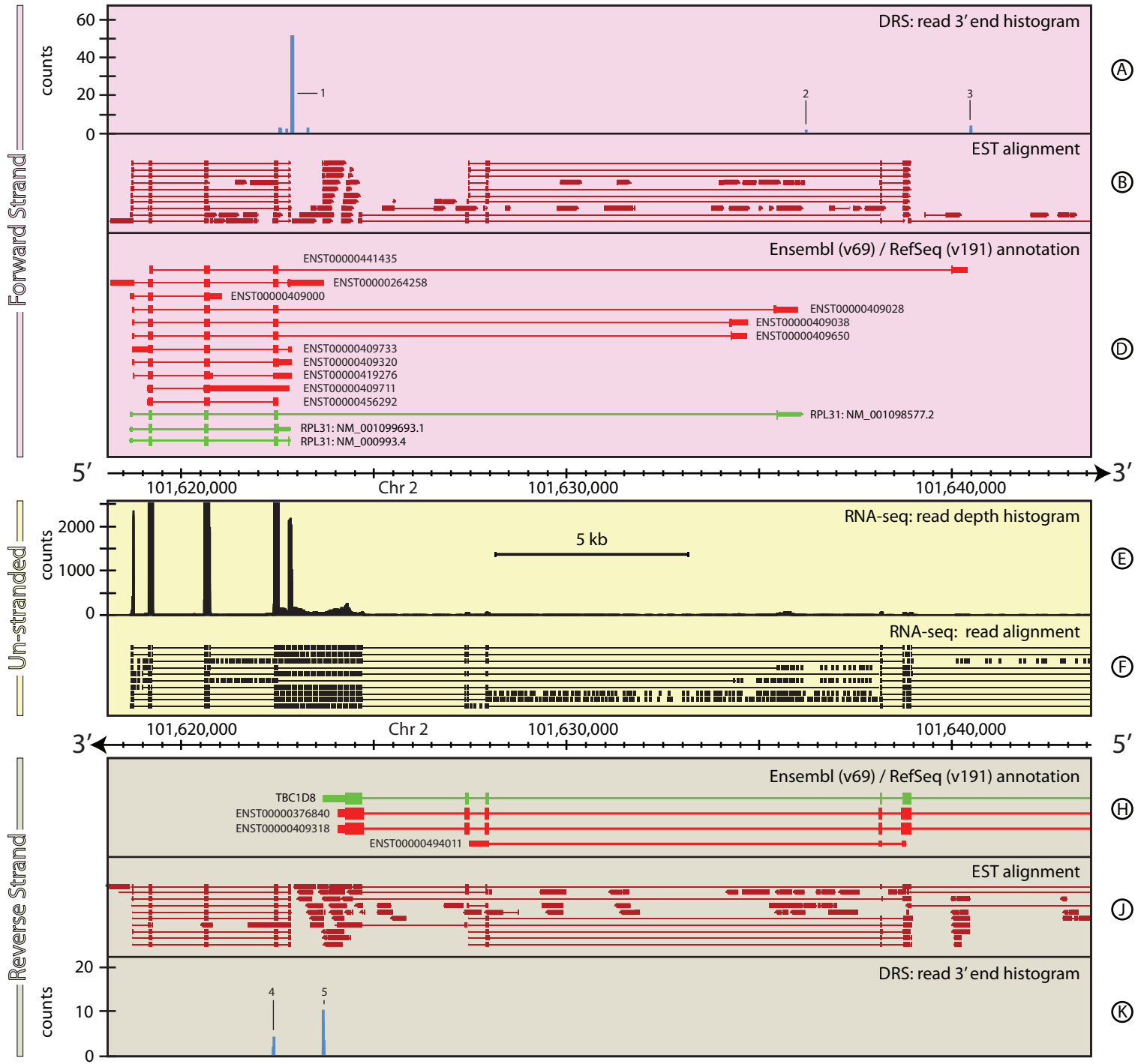

Figure 8.

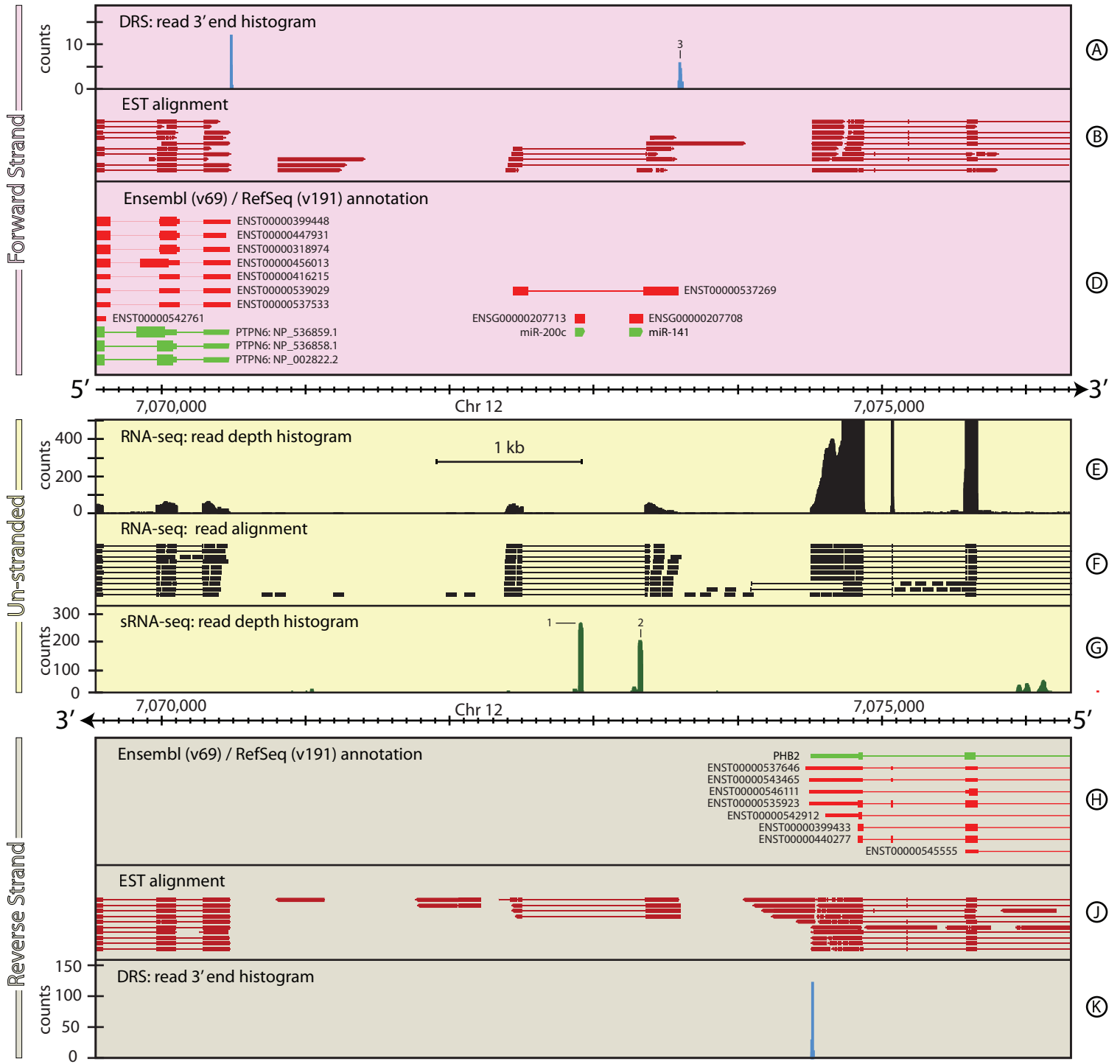

Figure 9.

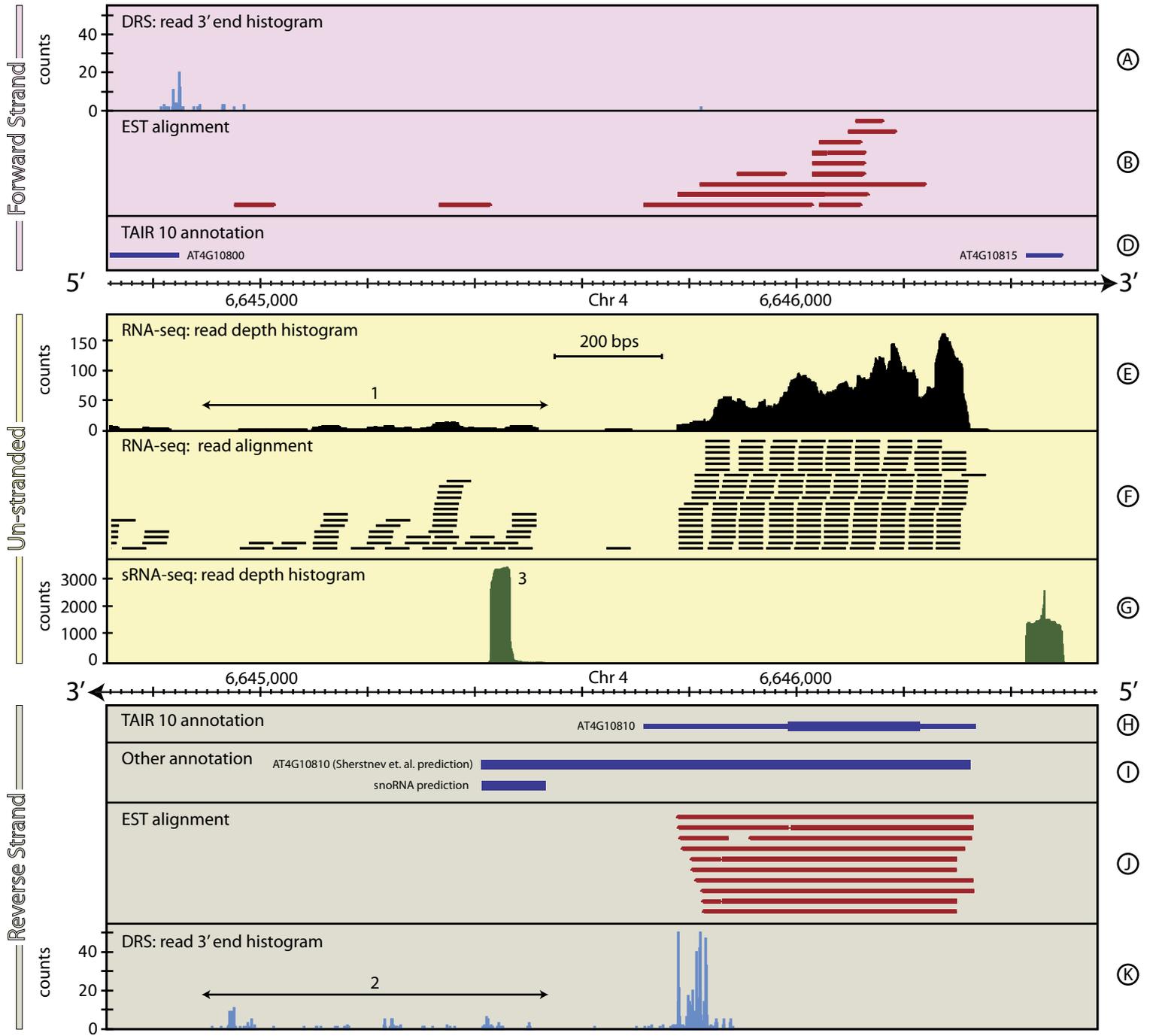